\documentclass[
  reprint,        
  aps,
  prb,            
  amsmath,amssymb,
  superscriptaddress
]{revtex4-2}

\usepackage{dcolumn}
\usepackage{amsmath}
\usepackage{amsfonts}
\usepackage{bm}
\usepackage{graphicx}
\usepackage{subcaption}

\usepackage{tikz}
\usetikzlibrary{positioning,arrows.meta,fit,calc}

\usepackage[colorlinks=true, allcolors=blue]{hyperref}

\usepackage{caption}
\usepackage{booktabs}

\graphicspath{{figs/}}

\begin{document}

\title{Spin Dynamics from Niu-Kleinman Adiabatic Approach and Slave Boson Mean Field Theory}
\author{Xuan Yang}
\affiliation{Beijing National Laboratory for Condensed Matter Physics and Institute of Physics,
	Chinese Academy of Sciences, Beijing 100190, China}
\affiliation{School of Physical Sciences, University of Chinese Academy of Sciences, Beijing 100190, China}

\author{Tianyang Xie}
\affiliation{Beijing National Laboratory for Condensed Matter Physics and Institute of Physics,
	Chinese Academy of Sciences, Beijing 100190, China}

\author{Shaohang Shi}
\affiliation{Beijing National Laboratory for Condensed Matter Physics and Institute of Physics,
	Chinese Academy of Sciences, Beijing 100190, China}
\affiliation{School of Physical Sciences, University of Chinese Academy of Sciences, Beijing 100190, China}

\author{Kun Jiang}
\email{jiangkun@iphy.ac.cn}
\affiliation{Beijing National Laboratory for Condensed Matter Physics and Institute of Physics,
	Chinese Academy of Sciences, Beijing 100190, China}
\affiliation{School of Physical Sciences, University of Chinese Academy of Sciences, Beijing 100190, China}

\author{Jiangping Hu}
\email{jphu@iphy.ac.cn}
\affiliation{New Cornerstone Science Laboratory, Beijing National Laboratory for Condensed Matter Physics and Institute of Physics,
	Chinese Academy of Sciences, Beijing 100190, China}

\begin{abstract}
    Spin-wave excitations provide a central probe of magnetic order and electronic correlations in strongly correlated materials. In this work, we develop an adiabatic theory of spin dynamics by combining the Niu–Kleinman formalism with Kotliar–Ruckenstein slave-boson theory (NK+KRSB). For each frozen spin configuration, the constrained slave-boson saddle point is solved self-consistently, allowing the Berry-curvature matrix and energy Hessian entering the linearized adiabatic equations of motion to be extracted directly. Applied to the half-filled single-orbital Hubbard model, the resulting spin-wave dispersion shows substantially improved agreement with determinant quantum Monte Carlo benchmarks compared with the random phase approximation and closely approaches results from the time-dependent Gutzwiller approximation. We further extend the method to a two-orbital model of La$_2$NiO$_4$, demonstrating its applicability to realistic multi-orbital correlated systems. Because the approach only requires saddle-point solutions near the magnetic ground state, it remains computationally efficient while incorporating strong-correlation effects beyond conventional weak-coupling descriptions, providing a practical framework for studying low-energy spin excitations in correlated quantum materials.
\end{abstract}

\maketitle

\section{Introduction}

Low-energy spin-wave dynamics provides a direct probe of magnetic order and electronic correlations in quantum materials~\cite{manousakis1991,auerbach1994}. It is closely connected to antiferromagnetism, Mott physics~\cite{imada1998,lee2006}, and possible magnetic mechanisms of superconducting pairing~\cite{scalapino2012,keimer2015}. A reliable and efficient description of spin excitations is therefore essential for understanding strongly correlated materials and for comparing theory with spectroscopic experiments such as inelastic neutron scattering and resonant inelastic x-ray scattering~\cite{tranquada2004,dean2015}.

Theoretical descriptions of spin dynamics are often developed from two complementary limits. In the strong coupling limit, the half-filled Hubbard model can be mapped to an effective Heisenberg model~\cite{anderson1950,macdonald$fractU$ExpansionHubbard1988}, whose magnetic excitations are commonly described by linear spin wave theory~\cite{holsteinFieldDependenceIntrinsic1939,manousakis1991,auerbach1994}. In the weak coupling limit, the system is treated as an itinerant electron problem, and Hartree-Fock (HF) theory combined with the random phase approximation (RPA) is widely used to compute the transverse spin susceptibility~\cite{schriefferDynamicSpinFluctuations1989,bulutKnightShiftsNuclearspinrelaxation1990,knolleMultiorbitalSpinSusceptibility2011}. These approaches are simple and physically transparent, but neither is generally reliable throughout the intermediate and strong correlation regimes relevant to many materials.

More accurate numerical methods provide important benchmarks but are difficult to apply broadly to realistic systems. Determinantal quantum Monte Carlo (DQMC)~\cite{blankenbeclerMonteCarloCalculations1981,hirschEfficientMonteCarlo1981,scalapinoMethodPerformingMonte1981,shaoProgressStochasticAnalytic2023,kungDopingEvolutionSpin2015,kungNumericallyExploring1D2D2017,huangNumericalEvidenceFluctuating2017,huangStripeOrderPerspective2018,pengDispersionDampingIntensity2018} can treat correlation effects in an unbiased way in sign-problem-free cases, but its computational cost remains high and analytic continuation is required to obtain real-frequency spectra ~\cite{sandvikStochasticMethodAnalytic1998,shaoProgressStochasticAnalytic2023}. Tensor network methods~\cite{jeckelmannDynamicalDensitymatrixRenormalizationgroup2002,kuhnerDynamicalCorrelationFunctions1999,daleyTimedependentDensitymatrixRenormalizationgroup2004,whiteRealTimeEvolutionUsing2004,haegemanTimeDependentVariationalPrinciple2011,haegemanUnifyingTimeEvolution2016,dargelAdaptiveLanczosvectorMethod2011,dargelLanczosAlgorithmMatrix2012,holznerChebyshevMatrixProduct2011,chiSpinExcitationSpectra2022} are also powerful, but in two dimensions the required bond dimension and contraction cost grow rapidly, making dynamical calculations especially demanding.

Based on Gutzwiller approximation or slave boson theory, there are two relatively cheap approaches to access the spin dynamics. One is Time-dependent Gutzwiller approximation (TDGA)~\cite{seiboldTimeDependentGutzwillerApproximation2001,seiboldInhomogeneousGutzwillerApproximation2003,seiboldTimedependentGutzwillerTheory2004,seiboldTimedependentGutzwillerTheory2008,bunemannLinearresponseDynamicsTimedependent2013}, which is similar to Time-dependent Hartree-Fock theory. It ultilize the time-dependent variational principle with trial wave function of Gutzwiller form. In the language of field theory, another one is to consider the Gaussian fluctuations of bosonic field at the saddle point~\cite{lavagnaFunctionalintegralApproachStrongly1990,liDynamicResponseFunctions1991,arrigoniCorrectContinuumLimit1995,zimmermannSpinChargeStructure1997,rieglerSlavebosonAnalysisTwodimensional2020,seufertBreakdownChargeHomogeneity2021a,mazumdarSlavebosonFormalismSuperconducting2026}. Both of them offer a better description of spin dynamics beyond HF+RPA, by incorporating the correlation effects through the Gutzwiller approximation.

An alternative efficient framework for spin dynamics was proposed by Niu and Kleinman, who developed an adiabatic theory of spin wave dynamics for \textit{ab initio} calculations~\cite{niuSpinWaveDynamicsReal1998,niuAdiabaticDynamicsLocal1999}. In this approach, slow collective spin dynamics are separated from high energy electronic degrees of freedom, and the spin wave equation of motion is constructed from the Berry curvature and the energy variation of frozen spin configurations. This adiabatic formulation is computationally attractive because it only requires the ground state corresponding to nearby constrained spin configurations. Previous implementations\cite{gebauerMagnonsRealMaterials2000,bylanderFeMagnonDispersion2000,renAdiabaticDynamicsCoupled2024,linAssessingAtomicMoment2025a,liuFullyFirstprinciplesApproach2026}, however, were mainly based on Hartree-Fock-level descriptions and therefore could not fully capture strong correlation effects. The formalism itself is more general: any method capable of determining the lowest energy state for a frozen spin configuration can, in principle, be combined with the Niu-Kleinman equation of motion. Related developments include reinterpretations of the adiabatic theory~\cite{qianSpinDynamicsTimeDependent2002}, density functional perturbation theory implementations~\cite{linAssessingAtomicMoment2025a,liuFullyFirstprinciplesApproach2026}, and extensions including coupled phonon dynamics~\cite{renAdiabaticDynamicsCoupled2024}.

In this work, we combine the Niu-Kleinman adiabatic spin dynamics formalism with Kotliar-Ruckenstein slave boson (KRSB) theory~\cite{kotliarNewFunctionalIntegral1986}. At the saddle point level, KRSB provides a field theoretic formulation closely related to the Gutzwiller approximation~\cite{kotliarNewFunctionalIntegral1986,Gutzwiller_PhysRevLett.10.159}: the auxiliary bosons encode local correlation effects, while the electron hopping is renormalized by correlation dependent quasiparticle factors. As a result, KRSB captures interaction induced quasiparticle renormalization and the Mott insulating transition~\cite{brinkmanApplicationGutzwillersVariational1970}, both of which are absent in Hartree-Fock theory. 
By using KRSB to solve the frozen spin configurations entering the Niu-Kleinman formalism, we obtain an efficient approach that incorporates important correlation effects into adiabatic spin wave dynamics.


The remainder of this paper is organized as follows. In Sec. II, we introduce the Niu-Kleinman adiabatic equation of motion. In Sec. III, we develop the method for the single orbital Hubbard model and compare the resulting spin wave dispersion with other approaches. In Sec. IV, we extend the formalism to a multi-orbital slave boson theory and apply it to a two-orbital model of La$_2$NiO$_4$. In Sec. V, we summarize the main results and discuss future directions.

\section{Niu-Kleinman Adiabatic Equation of Motion}

We first review the adiabatic equation of motion proposed by Niu and Kleinman~\cite{niuSpinWaveDynamicsReal1998,niuAdiabaticDynamicsLocal1999}. This formalism treats the local spin expectation values as slow collective coordinates and derives their dynamics from the time dependent variational principle. For a frozen spin configuration, the effective Lagrangian contains two ingredients: a Berry phase term associated with the evolution of the constrained electronic state, and the energy of that frozen spin configuration:
\begin{equation}
\begin{aligned}
    \mathcal{L} = \sum_{r,\alpha={x,y,z}} &\mathrm{i} \frac{\partial S_r^\alpha}{\partial t} \left\langle \Psi[\vec{S}(r)]\middle| \frac{\partial}{\partial S_r^\alpha} \middle| \Psi[\vec{S}(r)] \right\rangle  \\
    &- \left\langle\Psi[\vec{S}(r)]\middle| H \middle| \Psi[\vec{S}(r)]\right\rangle.
\end{aligned}
\end{equation}
$\vec{S}(r)=\left\langle \Psi[\vec{S}(r)]\middle|\hat{\vec{S}}_r\middle|\Psi[\vec{S}(r)]\right\rangle$ denotes the local spin expectation value in the constrained state. For each prescribed spin configuration $\vec{S}(r)$, $|\Psi[\vec{S}(r)]\rangle$ is taken to be the corresponding lowest energy electronic state. Varying the above Lagrangian with respect to the collective spin coordinates gives the equation of motion:
\begin{align}
    &-\sum_{\alpha',j'} \Omega_{jj'}^{\alpha\alpha'} S_{j'}^{\alpha'} + \frac{\partial E}{\partial S_j^\alpha} = 0, \label{NKEq}\\
    &\Omega_{jj'}^{\alpha\alpha'} = \frac{\partial}{\partial S_j^\alpha}\left\langle \Psi\middle| \frac{\mathrm{i}\partial}{\partial S_{j'}^{\alpha'}} \middle| \Psi \right\rangle - \frac{\partial}{\partial S_{j'}^{\alpha'}}\left\langle \Psi\middle| \frac{\mathrm{i}\partial}{\partial S_j^\alpha} \middle| \Psi \right\rangle.
\end{align}
For spin waves around a magnetic ground state, the transverse spin deviations are small. We therefore linearize Eq.~\ref{NKEq} around the equilibrium configuration:
\begin{equation}
    -\sum_{\alpha',j'} \Omega_{jj'}^{\alpha\alpha'} S_{j'}^{\alpha'} + \frac{\partial^2 E}{\partial S_j^\alpha \partial S_{j'}^{\alpha'}}S_{j'}^{\alpha'} = 0. \label{NKEqLinear}
\end{equation}
Equation~\ref{NKEqLinear} shows that the spin wave dynamics are determined by two quantities evaluated near the magnetic ground state: the Berry curvature of the constrained states and the energy curvature of the frozen spin configurations. Thus, the central computational task is to determine $|\Psi\rangle$ for spin configurations infinitesimally close to the ground state. Once these nearby constrained states are available, the adiabatic spin wave dynamics can be constructed independently of the particular method used to obtain them.


\section{Hubbard Model}

\subsection{Kotliar-Ruckenstein Slave Boson Theory}
Slave boson theory provides an efficient mean field framework for treating strongly correlated systems. It is computationally inexpensive while still capturing important correlation effects at a qualitative level. In the KRSB formulation, the physical local configurations are represented by auxiliary bosons together with pseudo fermions~\cite{kotliarNewFunctionalIntegral1986,liSpinrotationinvariantSlavebosonApproach1989,slave-boson-3,Jiang_PhysRevB.90.165135,Jiang_PhysRevLett.120.157205}. 
For the single orbital Hubbard model, the empty $\vert0\rangle$, singly occupied $\vert\uparrow\rangle,\vert\downarrow\rangle$, and doubly occupied  $\vert\uparrow\downarrow\rangle$ configurations are assigned separate bosonic operators:
$e, p_\uparrow, p_\downarrow, d$.
The physical electron operator at site $i$ with spin $\sigma$ is written as a pseudo fermion dressed by a local renormalization factor,
\begin{equation}
    c_{i\sigma}^\dagger = z_{i\sigma}^\dagger f_{i\sigma}^\dagger,
\end{equation}
where $ z_{i\sigma}^\dagger =L_i^{-\frac12} \left(p_{i\sigma}^\dagger e_i+d_i^\dagger p_{i\bar\sigma}\right) R_i^{-\frac12} $
and $L_i=1-d_i^\dagger d_i-p_{i\sigma}^\dagger p_{i\sigma}$, $R_i=1-e_i^\dagger e_i-p_{i\bar\sigma}^\dagger p_{i\bar\sigma}$.
The enlarged Hilbert space is constrained by
\begin{align}
    &e_i^\dagger e_i+\sum_\sigma p_{i\sigma}^\dagger p_{i\sigma}+d_i^\dagger d_i=1,\\
    &f_{i\sigma}^\dagger f_{i\sigma}=p_{i\sigma}^\dagger p_{i\sigma}+d_i^\dagger d_i.
\end{align}
In this representation, the local Hubbard interaction term becomes simple, while the hopping term acquires correlation dependent renormalization factors but remains bilinear in the pseudo fermions:
\begin{equation}
    H =
    \sum_{\langle ij\rangle,\sigma}
    -f_{i\sigma}^\dagger z_{i\sigma}^* t_{ij} z_{j\sigma} f_{j\sigma}
    +
    U\sum_i d_i^\dagger d_i.
\end{equation}
At the saddle point level, this construction gives a mean field theory equivalent to the Gutzwiller approximation.

The original KRSB representation is most naturally written in a spin quantization axis fixed by the magnetic order. For the present problem, however, we need to describe transverse spin fluctuations and locally rotated spin configurations. It is therefore useful to adopt the rotationally invariant Kotliar-Ruckenstein slave boson (RI-KRSB) formalism~\cite{liSpinrotationinvariantSlavebosonApproach1989}, in which the spin direction can vary continuously while spin rotation symmetry remains explicit.

In the RI-KRSB formulation, the only modification lies in the singly occupied sector.  The $p_\sigma$ bosons are generalized by a slave boson matrix written in terms of the Pauli matrices $\tau$:
\begin{equation}
    \mathbf{p}= \frac1{\sqrt2} (p_0 \tau^0 + \vec{p}\cdot\vec{\tau})=\frac{1}{\sqrt2}
    \begin{pmatrix}
    	p_0 + p_z & p_x - i p_y \\
    	p_x + i p_y & p_0 - p_z
    \end{pmatrix}
\end{equation}
This parametrization makes spin rotation symmetry explicit. The singly occupied physical states and the pseudo fermion $f$ both carry spin $\frac{1}{2}$. 
Then, the constraints must be imposed to project back to the physical subspace:
\begin{align}
    &e^\dagger e + p_0^\dagger p_0 + p_x^\dagger p_x + p_y^\dagger p_y + p_z^\dagger p_z + d^\dagger d = 1, \\
    &\sum_{\sigma}f_\sigma^\dagger f_\sigma = \sum_{\mu\in \{0,x,y,z\}} p_\mu^\dagger p_\mu + 2d^\dagger d, \\
    &\sum_{\sigma,\sigma'}f_{\sigma'}^\dagger \vec{\tau}_{\sigma,\sigma'} f_\sigma = p_0^\dagger \vec{p} + \vec{p}^\dagger p_0 + \mathrm{i} \vec{p}^\dagger \times \vec{p}.
\end{align}

Within the constrained Hilbert space, the physical electron operators are represented by pseudo fermions dressed by slave boson renormalization matrices:
\begin{align}
    &c_{i\sigma}^\dagger = \sum_{\sigma'}Z_{i\sigma\sigma'}^\dagger f_{i\sigma'}^\dagger, \\ 
    Z_i = & \mathbf{L}_i^{-\frac12} \left(e_i^\dagger \mathbf{p}_i  + \tilde{\mathbf{p}}_i^\dagger d_i\right) \mathbf{R}_i^{-\frac12}, \\
    \mathbf{L}_i =& (1-d_i^\dagger d_i)\tau^0 - \mathbf{p}_i^\dagger \mathbf{p}_i, \\
    \mathbf{R}_i =& (1-e_i^\dagger e_i)\tau^0 -\tilde{\mathbf{p}}_i^\dagger \tilde{\mathbf{p}}_i,
\end{align}
where $\mathbf{p}_i = \frac1{\sqrt2} (p_{0,i} \tau^0 + \vec{p}_i\cdot\vec{\tau})$ and $\tilde{\mathbf{p}}_i =\hat{\mathbf{T}} \mathbf{p}_i\hat{\mathbf{T}}^{-1} = \frac1{\sqrt2} (p_{0,i} \tau^0 - \vec{p}_i\cdot\vec{\tau})$.

At the saddle point, the bosons are treated as condensed classical fields. The Hubbard model then becomes an effective bilinear Hamiltonian for the pseudo fermions, with the hopping renormalized by the slave boson matrices:
\begin{equation}
    \begin{aligned}
        H &= \sum_{\langle ij\rangle,\sigma} -c_{i\sigma}^\dagger t_{ij} c_{j\sigma} + U \sum_i n_{i\uparrow} n_{i\downarrow} \\
        &= \sum_{\langle ij\rangle,\sigma,\sigma',\sigma''} -f_{i\sigma}^\dagger Z_{i,\sigma\sigma'}^* t_{ij} Z_{j,\sigma'\sigma''}^T f_{j,\sigma''} + U\sum_{i} d_i^\dagger d_i.
    \end{aligned}
\end{equation}

Having introduced the general RI-KRSB representation, we now specify the magnetic configurations used in the adiabatic spin wave calculation. This ansatz determines how the slave bosons and Lagrange multipliers are parametrized for each constrained state. We take the ordered ground state to be a two sublattice antiferromagnet with staggered magnetization along the $z$ direction, and describe low energy spin waves as slow transverse rotations of this order. The collective variables entering the Niu-Kleinman equation are therefore the in plane spin components $\{S_\alpha^x, S_\alpha^y\}$, where $\alpha=A,B$ labels the two sublattices. These transverse components are fixed when solving the saddle point equations. The remaining amplitude related degrees of freedom are treated as fast variables and are allowed to relax for each fixed transverse spin configuration. In particular, $d_\alpha$ and $p_\alpha^z$ are optimized independently on the two sublattices. Since different spin wave momenta decouple at the linearized level, we consider one wavevector $\vec{q}$ at a time. The corresponding transverse spin texture is written as
\begin{equation}
    \begin{aligned}
        S_{\alpha,i}^x + \mathrm{i}S_{\alpha,i}^y &= (S_{\alpha,q}^x + \mathrm{i} S_{\alpha,q}^y) \mathrm{e}^{\mathrm{i}\vec{q}\cdot\vec{r}_i} \\
        &= S_{\alpha,q}^{\parallel} \mathrm{e}^{\mathrm{i} \varphi_\alpha} \mathrm{e}^{\mathrm{i}\vec{q}\cdot\vec{r}_i}
    \end{aligned}
\end{equation}

The effective free energy is obtained by integrating out the pseudo fermions. It depends on the slave bosons and Lagrange multipliers, and the saddle point solution is found by minimizing this free energy:

\begin{equation}
    \begin{aligned}
        S_{eff}&[d,p_0,\vec{p},\lambda] = -\frac1{\beta}\sum_k \mathrm{tr}\ln\left(1+e^{-\beta h_k}\right) + \frac N2 U \sum_\alpha d_\alpha^2  \\
        &+ \frac N2\sum_\alpha\left[ \lambda_{\alpha 0}(p_{\alpha 0}^2 + \vec{p}_{\alpha}^2 + 2d_\alpha^2) + \vec\lambda_\alpha \cdot (2p_{\alpha 0}\vec{p}_\alpha) \right].
    \end{aligned}
\end{equation}
Here $h_k$ is the momentum space mean field Hamiltonian, which depends on the slave bosons and Lagrange multipliers. The boson $e$ has been eliminated using the first constraint. The subscript $\alpha$ denotes the sublattice index, appropriate for the antiferromagnetic state considered here.

For the free energy evaluation, it is useful to decompose the hopping term into independent $k$ sectors in the Brillouin zone of the magnetic unit cell. This step is important for making the calculation efficient. The Fourier transformed mean field Hamiltonian takes the form
\begin{align}
    H_{MF} &= \sum_{k} \psi_k^\dagger h_k \psi_k, \\
    \psi_k &= (f_{A,k,\downarrow}, f_{B,k,\downarrow}, f_{A,k+q,\uparrow}, f_{B,k+q,\uparrow})^T, \\
    h_k &= \sum_{\delta} (-t\exp(\mathrm{i}(\vec k+\frac {\vec q}2)\cdot\vec{\delta}))\sigma^+M(q,\delta) + h.c.  \notag\\
    &+ \left( \frac{\lambda_A^z+\lambda_B^z}2\sigma^0 + \frac{\lambda_A^z-\lambda_B^z}2\sigma^z \right)\tau^z  \notag\\
    &+ \left( \frac{\lambda_A^++\lambda_B^+}2\sigma^0 + \frac{\lambda_A^+-\lambda_B^+}2\sigma^z \right)\tau^+  \notag\\
    &+ \left( \frac{\lambda_A^-+\lambda_B^-}2\sigma^0 + \frac{\lambda_A^--\lambda_B^-}2\sigma^z \right)\tau^-, \label{h_kSingleOrb}\\
    M(q,\delta) &= U_A^* Z_A^{dia} U_A^T U_z^*(\vec{q}\cdot\vec{\delta}) U_B^* Z_B^{dia} U_B^T.
\end{align}
$\delta$ is the nearest neighbor vector; $\sigma$ denotes the Pauli matrices in sublattice space; $\tau$ denotes the Pauli matrices in spin space; $Z^{dia}$ is the diagonal renormalization matrix obtained when only the $z$ component of the $p$ boson is retained; $U_z(q\cdot\delta)$ is the spin rotation matrix for a rotation angle $q\cdot\delta$ around the $z$ axis; and $U_A,U_B$ are the spin rotation matrices for sublattices A and B, respectively, determined by the spin polarization on each sublattice. The details of the Fourier transformation are omitted for brevity.

\subsection{Parameterization of Slave Bosons And Saddle Point Equation}

In order to obtain the Niu-Kleinman adiabatic equation, we need the perturbed saddle point for a fixed in plane spin configuration $S^x$, $S^y$. In RI-KRSB theory, the spin configuration $\vec{S}$ is determined by the $p$ bosons. For given $S^x$ and $S^y$, we have
\begin{equation}
\begin{aligned}
    p_0p^x &= S^x \\
    p_0p^y &= -S^y \\
    p_0^2 + \vec{p}^2 + e^2 + d^2 &=1 \\
    p_0^2 + \vec{p}^2 + 2d^2 &= n.
\end{aligned}
\end{equation}
Using the above constraints, the other bosons can be expressed in terms of $d$, $p^z$, $n$, $S^x$, and $S^y$.
Substituting these bosons into the action and keeping only $d$ and $p^z$ as independent variables, we obtain the effective action $S_{eff}'(d,p^z)$ for given $S^x$ and $S^y$.

Because the ground state is antiferromagnetic, the bosons on the two sublattices $\alpha$ are different. The saddle point equations with respect to $p^z$ and $d$ are
\begin{align}
    \frac{1}{N}\sum_{k,m} n_F(\epsilon_k^m)\frac{\partial\epsilon_k^m}{\partial p_\alpha^z} - 2\lambda_\alpha^z(\frac{\partial p_{\alpha 0}}{\partial p_\alpha^z} p_\alpha^z + p_{\alpha 0}) &=0 \\
    \frac{1}{N}\sum_{k,m} n_F(\epsilon_k^m)\frac{\partial\epsilon_k^m}{\partial d_\alpha} + 2 U d_\alpha - 2\lambda_\alpha^z \frac{\partial p_{\alpha 0}}{\partial d_\alpha}p_\alpha^z &=0.
\end{align}
$\epsilon_k^m$ is the eigenvalue of $h_k$. After solving the above equations together with the constraint equations, we obtain $p_\alpha^z,d_\alpha,\lambda_{\alpha 0},\vec{\lambda}_\alpha$, which parametrize the state we need within the slave boson mean field formalism.

\subsection{Computation Process}

\begin{figure}[htbp]
\centering
\begin{tikzpicture}[
    >=Latex,
    font=\footnotesize,
    node distance=6mm,
    mainbox/.style={
        draw,
        rounded corners,
        thick,
        align=center,
        text width=0.82\columnwidth,
        minimum height=9mm,
        inner sep=4pt,
        fill=blue!5
    },
    branchbox/.style={
        draw,
        rounded corners,
        thick,
        align=center,
        text width=0.36\columnwidth,
        minimum height=10mm,
        inner sep=4pt,
        fill=green!8
    },
    groupbox/.style={
        draw,
        dashed,
        thick,
        rounded corners,
        inner sep=5pt
    },
    flow/.style={->, thick}
]

\node[mainbox] (q) {Fix a spiral wavevector $\vec{q}$};

\node[mainbox, below=of q] (config) {Prepare a family of magnetic configurations around the ground state:\\
for the variables $\{S_{Aq}^x,S_{Aq}^y,S_{Bq}^x,S_{Bq}^y\}$, construct the grid points required for numerical differentiation.};

\node[mainbox, below=of config] (saddle) {Solve the saddle point solutions for all the grid points};

\node[groupbox, fit=(config)(saddle),
      label={[font=\footnotesize]above:Solve a family of saddle point equations}] (family) {};

\coordinate (split) at ($(saddle.south)+(0,-8mm)$);

\node[branchbox, anchor=north east] (berry) at ($(split)+(-2mm,0)$)
{Extract the Berry curvature from the saddle point solutions};

\node[branchbox, anchor=north west] (hessian) at ($(split)+(2mm,0)$)
{Extract the energy Hessian matrix from total energy variations};

\node[mainbox, below=24mm of saddle] (omega) {Combine both matrices to obtain the equation of motion at this $\vec{q}$};

\draw[flow] (q) -- (config);
\draw[flow] (config) -- (saddle);
\draw[flow] (saddle) -- (berry);
\draw[flow] (saddle) -- (hessian);
\draw[flow] (berry) -- (omega);
\draw[flow] (hessian) -- (omega);

\end{tikzpicture}
\caption{Workflow for computing the spin wave energy at a fixed spiral wavevector $\vec{q}$. The central step is to obtain a family of saddle point solutions for the same $\vec{q}$, including the reference state and nearby perturbed states, from which the Berry curvature matrix and the energy Hessian are constructed.}
\label{fig:fixed-q-workflow}
\end{figure}
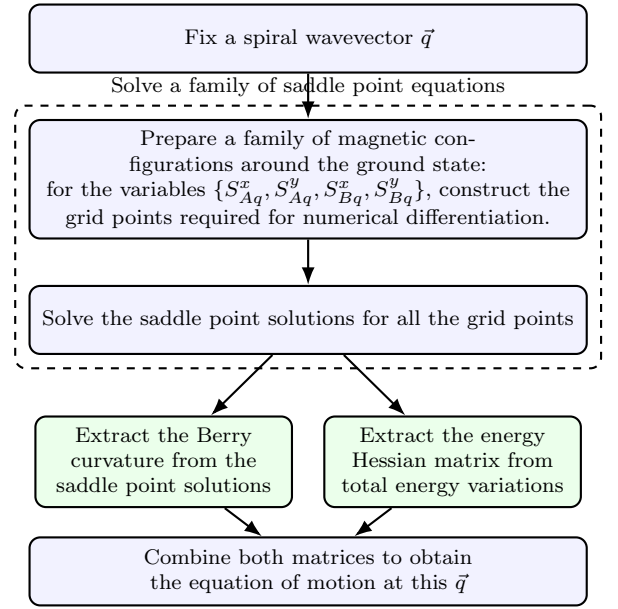

For each spin wave momentum $\vec{q}$, the numerical workflow is summarized in Fig.~\ref{fig:fixed-q-workflow}. We first construct a set of transverse spin configurations near the magnetic ground state. These configurations are finite difference points in the space of collective variables $\{S_{Aq}^x,S_{Aq}^y,S_{Bq}^x,S_{Bq}^y\}$. For each point, the transverse spin components are fixed, while the remaining slave bosons and Lagrange multipliers are optimized by solving the saddle point equations. The resulting constrained saddle point states are then used to evaluate the Berry curvature matrix and the energy Hessian. Combining these two matrices gives the linearized Niu-Kleinman equation of motion at the chosen $\vec{q}$.

The most expensive part of this procedure is solving the family of nearby saddle point equations. In practice, this cost is moderate because all constrained configurations are close to the ground state, so the ground state saddle point provides an efficient initial guess. Moreover, the saddle point calculations for different finite difference points are independent and can be parallelized straightforwardly.

\subsection{Results}

\begin{figure}[htbp]
    \includegraphics[width=\linewidth]{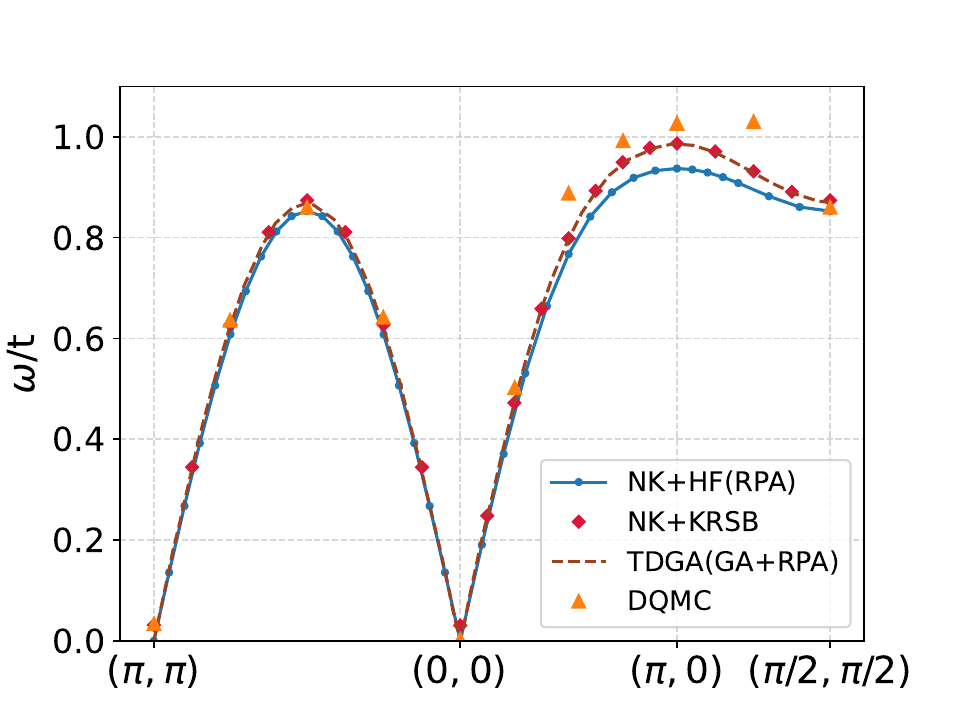}
    \caption{Spin wave dispersion of the half filled single orbital Hubbard model obtained from different methods. The parameters are $U=8.0$ and $n=1.0$. The method developed in this work is denoted by NK+KRSB, and the calculation is performed on $80\times80$ magnetic unit cells. The triangle markers show DQMC results on an $8\times8$ lattice, which has no sign problem at half filling. NK+KRSB, similar to TDGA, agrees better with the DQMC results than RPA does, especially around $\vec{q}=(\pi,0)$.}
    \label{fig:SingleOrbDispersion}
\end{figure}

We first test the method on the half filled single orbital Hubbard model on a square lattice. For the interaction strength considered here, $U=8$, the ground state is a N\'eel antiferromagnet with ordering vector $(\pi,\pi)$. Figure~\ref{fig:SingleOrbDispersion} shows the calculated spin wave dispersion along high symmetry momentum cuts, together with results obtained from other approaches.

We denote the method developed in this paper by NK+KRSB, meaning the combination of the Niu-Kleinman adiabatic equation and Kotliar-Ruckenstein slave boson theory. For comparison, NK+HF denotes the same adiabatic construction using Hartree-Fock mean field theory. The NK+HF dispersion nearly coincides with the RPA result, indicating that the Niu-Kleinman formalism combined with a Hartree-Fock constrained state reproduces the usual weak coupling transverse spin dynamics in the low energy regime. In contrast, NK+KRSB substantially improves the dispersion relative to RPA and agrees much better with DQMC, especially near $\vec{q}=(\pi,0)$. The DQMC results are obtained by calculating the imaginary time spin correlation function and then performing analytic continuation using stochastic analytic continuation (SAC). The parameters of DQMC calculation are $n=1.0$, $L_x=8$, $L_y=8$, and $\beta=8t$. The SAC parameter is $1.2^{10}$. The TDGA results are adapted from Ref.~\cite{lorenzanaSumRulesMissing2005}.

The improvement of NK+KRSB over NK+HF originates from the slave boson renormalization of the kinetic term in the strong interaction regime. The NK+KRSB dispersion is also close to the TDGA result, consistent with the relation between KRSB saddle point theory and the Gutzwiller approximation. This agreement suggests that the present adiabatic construction captures the same low energy correlation induced renormalization of spin dynamics at the slave boson mean field level. At the same time, NK+KRSB is computationally simple: instead of solving a full time dependent variational problem, one only needs constrained saddle point solutions infinitesimally close to the magnetic ground state.

\section{Multi-Orbital Hubbard Model}

\subsection{Two Orbital Slave Boson Theory}
We now generalize the slave boson formulation to the multi-orbital case. 
Here, we take a two-orbital model as one example. 
For a two-orbital model, we consider the local Kanamori interaction Hamiltonian
\begin{equation}
    \begin{aligned}
	H_{loc} &= \sum_{o=\{1,2\}} U n_{o,\uparrow}n_{o,\downarrow} + \sum_{\sigma,\sigma'}(U-2J_H) n_{1,\sigma}n_{2,\sigma'}  \\
	&- J_H \sum_{\sigma} n_{1,\sigma}n_{2,\sigma} + J_H\sum_\sigma d_{1,\sigma}^\dagger d_{2,\bar\sigma}^\dagger d_{1,\bar\sigma} d_{2,\sigma} \\
	&+J_H \left( d_{1,\uparrow}^\dagger d_{1,\downarrow}^\dagger d_{2,\downarrow} d_{2,\uparrow} + d_{2,\uparrow}^\dagger d_{2,\downarrow}^\dagger d_{1,\downarrow} d_{1,\uparrow} \right).
    \end{aligned}
\end{equation}
The main difficulty in constructing a multi-orbital slave boson theory is to keep both the local interaction terms and the constraint terms manageable. This can be achieved by separating the physical local states from the quasiparticle Fock states~\cite{lechermannRotationallyInvariantSlaveboson2007,Gehbard_PhysRevB.57.6896,Sigrist_PhysRevLett.92.216402,Dai_2006cond.mat.11075D,Deng_2008,Dai_PhysRevB.79.075114,Hellsing_PhysRevB.85.035133,Dai_EPB,Dai_PhysRevB.86.035150,Dai_PhysRevLett.102.126401}.

In the rotationally invariant slave boson formalism \cite{lechermannRotationallyInvariantSlaveboson2007}, a slave boson is introduced for each pair of physical and quasiparticle states. A physical state in the enlarged Hilbert space can then be represented as
\begin{equation}
\left| \underline{A} \right\rangle \equiv \frac1{\sqrt{D_A}} \sum_n \phi_{An}^\dagger |vac\rangle \otimes |n\rangle_f.
\end{equation}
The physical subspace is selected by the constraints
\begin{align}
    \sum_{An} \phi^\dagger_{An}\phi_{An} &= 1 \\
    \sum_A \sum_{nn'} \phi^\dagger_{An'} \phi_{An} \langle n | f_\alpha^\dagger f_{\alpha'} | n'\rangle &= f_\alpha^\dagger f_{\alpha'}, \forall \alpha\alpha'.
\end{align}
Within this constrained subspace, the physical electron operator is represented by a quasiparticle operator dressed by a slave boson renormalization matrix:
\begin{align}
    \underbar{d}_\alpha^\dagger &= \hat{R}[\phi]_{\alpha\beta}^* f^\dagger_\beta, \\
    \hat{R}[\phi]_{\alpha\beta}^* &= \sum_{AB,nm,\gamma}\langle A|d_\alpha^\dagger|B\rangle\langle n|f_\gamma^\dagger|m\rangle \phi_{An}^\dagger \phi_{Bm} M_{\gamma\beta},\\
    M_{\gamma\beta} &\equiv \left\{ \left[\frac12\left(\hat{\Delta}^{p}\hat{\Delta}^{h} + \hat{\Delta}^{h}\hat{\Delta}^{p}\right)\right]^{-1/2} \right\}_{\gamma\beta}, \label{M_definition}\\
    \hat{\Delta}^{p} &\equiv \sum_{Anm}\phi_{An}^\dagger\phi_{Am}\langle m|f_\alpha^\dagger f_\beta|n\rangle, \\
    \hat{\Delta}^{h} &\equiv \sum_{Anm}\phi_{An}^\dagger\phi_{Am}\langle m|f_\beta f_\alpha^\dagger|n\rangle.
\end{align}

With this construction, the kinetic term remains bilinear in the quasiparticles, while the local interaction term is expressed directly in terms of the slave bosons. The mean field theory is obtained by condensing the $\phi$ bosons. The resulting saddle point problem has the same structure as in the single orbital case, but involves a much larger set of bosonic amplitudes and Lagrange multipliers:
\begin{equation}
    \begin{aligned}
        H_{MF} =& \sum_{ij} f_i^\dagger\,R_i^\dagger\,t_{ij}\,R_j\,f_j   \\
        &+ \sum_{i,ABn}\langle A|H_{loc}|B\rangle \phi_{An}^*\phi_{Bn}  \\
        &+ \sum_{i}\lambda_{i,0}(\sum_{An}\phi_{i,An}^*\phi_{i,An}-1)  \\
        &+ \sum_{i,\alpha\alpha'} \lambda_{i,\alpha\alpha'} \big(\sum_{Anm} \phi_{i,An}^* \phi_{i,Am} \langle m|f_\alpha^\dagger f_{\alpha'}|n\rangle \\
        &\qquad - f_{i,\alpha}^\dagger f_{i,\alpha'}\big).
    \end{aligned} \label{HMF_twoorb_realspace}
\end{equation}

\subsection{Parameterization of Slave Bosons}
\label{ParameterizationTwoOrb}

Taking both $|A\rangle$ and $|n\rangle$ as Fock basis, there are 16 diagonal slave bosons, 120 off diagonal slave bosons, and 16 Lagrange multipliers in the two orbital case, which is too large a number to solve for and not easy to interpret. Fortunately, we are interested only in the spin dynamics of the system, so we can parametrize the slave bosons according to the spin configuration. Since our focus is spin dynamics, we assume that the slave bosons are parametrized by a spin space rotation of the slave bosons for spin polarization along the $z$ axis, which are almost entirely diagonal. The rotation angle is determined by the spin polarization on the site.

For example, for a site with spin polarization $\mathbf{S}=\left(S_x,S_y,S_z\right)$,
\begin{align}
    \phi_{An} &= \sum_{A'} U_{AA'}(\theta,\varphi) \bar\phi_{A'n} \\
    U(\theta,\varphi) &= \exp\left(-\mathrm{i}\frac\varphi2\sigma^z\right)\exp\left(-\mathrm{i}\frac\theta2\sigma^y\right) \\
    \tan\varphi = \frac{S_y}{S_x}, &\qquad \tan\theta = \frac{\sqrt{S_x^2+S_y^2}}{S_z}.
\end{align}
The $\bar\phi_{A'n}$ are the slave bosons for the state with spin polarization along the $z$ axis and are almost diagonal. $U$ is spin rotation matrix in the local Hilbert space written in basis $A$. After this rotation, the spin polarization of the state is
\begin{align}
    S^a(\phi) &= \sum_{ABn} \langle A | \hat S^a | B\rangle \phi_{An}^* \phi_{Bn} \notag \\
    &= \sum_{A'B'n} \langle A' | \hat U^\dagger \hat S^a \hat U | B' \rangle \bar\phi_{A'n}^* \bar\phi_{B'n} \notag \\
    &= O^{ab} \sum_{A'B'n} \langle A' | \hat S^b | B' \rangle \bar\phi_{A'n}^* \bar\phi_{B'n} \notag \\
    &= O^{ab} S^b(\bar\phi).
\end{align}
$O$ is $O(3)$ rotation matrix for spin. It rotates the spin from the $z$ axis to the direction determined by $\theta$ and $\varphi$. 

This parametrization has two advantages. First, the number of parameters is greatly reduced and becomes almost the same as in the usual AFM ansatz. Second, rotating only the left index of $\phi$ does not change the local interaction term ($H_{loc}$ is spin rotation invariant) or the constraint term.
\begin{align}
    \sum_{ABn} \langle A|H_{loc}&|B\rangle \phi_{An}^*\phi_{Bn} \notag\\
    &= \sum_{ABn,A'B'} \langle A|H_{loc}|B\rangle U_{AA'}^* \bar{\phi}_{A'n}^* U_{BB'} \bar{\phi}_{B'n} \notag \\
    &= \sum_{A'B'n} \langle A'| \hat U^\dagger H_{loc} \hat U |B'\rangle \bar{\phi}_{A'n}^* \bar{\phi}_{B'n}, \\
    \sum_{An}\phi_{An}^*\phi_{An} &= \sum_{An,BB'} U_{AB}^* \bar\phi_{Bn}^* U_{AB'} \bar\phi_{B'n} \notag\\
    &= \sum_{Bn} \bar\phi_{Bn}^* \bar\phi_{Bn}, \\
    \sum_{Anm} \phi_{An}^* \phi_{Am} &\langle m|f_\alpha^\dagger f_{\alpha'}|n\rangle \notag\\
    &= \sum_{Anm,BB'} U_{AB}^* \bar\phi_{Bn}^* U_{AB'} \bar\phi_{B'm} \langle m|f_\alpha^\dagger f_{\alpha'}|n\rangle \notag \\
    &= \sum_{Anm} \bar\phi_{An}^* \bar\phi_{Am} \langle m|f_\alpha^\dagger f_{\alpha'}|n\rangle.
\end{align}

The rotation matrix $U$ enters only the hopping terms. As in the single orbital case, the full Hamiltonian can be Fourier transformed into momentum space, and the $f_k$ with different $k$ are decoupled. After some tedious derivation, the mean field Hamiltonian becomes block diagonal in momentum space, and $H_k$ is
\begin{equation}
    \begin{aligned}
        H_{kin} =& \sum_{k,\delta} f_{a^0_\delta k}^\dagger \bar{R}_{a^0_\delta}^\dagger \tilde{u}_{a^0_\delta}^T \tilde{u}_z^*(q\cdot\delta) \tilde{u}_{a^1_\delta}^* \bar{R}_{a^1_\delta} f_{a^1_\delta k} \\
        \\& + \sum_{a=\{A,B\}}f_{ak}^\dagger \Lambda_a f_{ak}, \\
        \tilde{u}_{a} =& \tilde{u}_z(\varphi_a) \tilde{u}_y(\theta_a).
    \end{aligned}\label{h_kTwoOrb}
\end{equation}
$f_{ak}$ is a four dimensional spinor with four spin orbital components, $a$ and $k$ are sublattice and momentum indices respectively; $t_\delta$ is a $4\times4$ hopping matrix that contains no spin flip terms; $a_{\delta}^0$ and $a_{\delta}^1$ are the starting and ending sublattice indices of hopping $\delta$; $\bar{R}_a$ is the renormalization matrix for site $a$, which is a function of the $\bar\phi$ bosons; $\tilde{u}_y$ and $\tilde{u}_z$ are spin rotation matrices in spin orbital space determined by the spin configuration; and $\Lambda_a$ is the Lagrange multiplier matrix for site $a$. The derivation of this equation is in the appendix.

After deriving the mean field Hamiltonian, the saddle point equations can be obtained. All calculation steps are almost the same as in the single orbital case, except for the Berry curvature calculation discussed in the next subsection. One point worth mentioning is that the high energy degrees of freedom here, namely all elements of $\bar\phi$, are much more numerous than in the single orbital case that contains only $d$ and $p^z$.

\subsection{Berry Curvature Calculation for Two Orbital Case}

The evaluation of the Berry curvature in the two-orbital formulation requires additional care. In particular, it cannot be computed directly from the Hamiltonian in Eq.~\ref{h_kTwoOrb}, because the pseudo-fermion states are defined in a different gauge choice from that used in the single-orbital formulation. As a result, the naive expression for the Berry curvature becomes gauge dependent. The difference between the gauge conventions in the single- and two-orbital formalisms is discussed in Appendix~\ref{DiffOfBerry}.

The Berry curvature should instead be evaluated using the following procedure. After solving a family of saddle point solutions around the ground state, The Berry curvature is calculated by the same finite difference formula as in the single orbital case. However, the Hamiltonian $H_k'$ when calculating the Berry curvature should be
\begin{equation}
    \begin{aligned}
        H_k' &= \sum_{k,\delta} f_{a^0_\delta k}^\dagger \tilde{u}_{a^0_\delta}^* \bar{R}_{a^0_\delta}^\dagger \tilde{u}_{a^0_\delta}^T \tilde{u}_z^*(q\cdot\delta) t_\delta \tilde{u}_{a^1_\delta}^* \bar{R}_{a^1_\delta} \tilde{u}_{a^1_\delta}^T f_{a^1_\delta k} \\
        \\& + \sum_{a=\{A,B\}}f_{ak}^\dagger \Lambda_a f_{ak}.
    \end{aligned}\label{hk_TwoOrbBerry}
\end{equation}
As discussed in the Appendix~\ref{DiffOfBerry}, $\phi$ is required to be Hermitian. Thus, we rotate $\phi=U\bar{\phi}$ on the right as well, which is $U\bar{\phi}U^\dagger$. A small trick worth mentioning is that we ignore the coordinate dependent rotation $U_z(\vec{q}\cdot \vec{r}_i)$ on the right rotation matrix to make the Fourier transformation simpler, otherwise it's hard to obtain the momentum separated Hamiltonian Eq.~\ref{hk_TwoOrbBerry}. It doesn't change the Berry curvature because $U_z(\vec{q}\cdot \vec{r}_i)$ is independent of the variables $\{S_{Aq}^x,S_{Aq}^y,S_{Bq}^x,S_{Bq}^y\}$.

\subsection{Application to \texorpdfstring{La$_2$NiO$_4$}{La2NiO4}}

\begin{figure}[htbp]
    \centering
    \includegraphics[width=\linewidth]{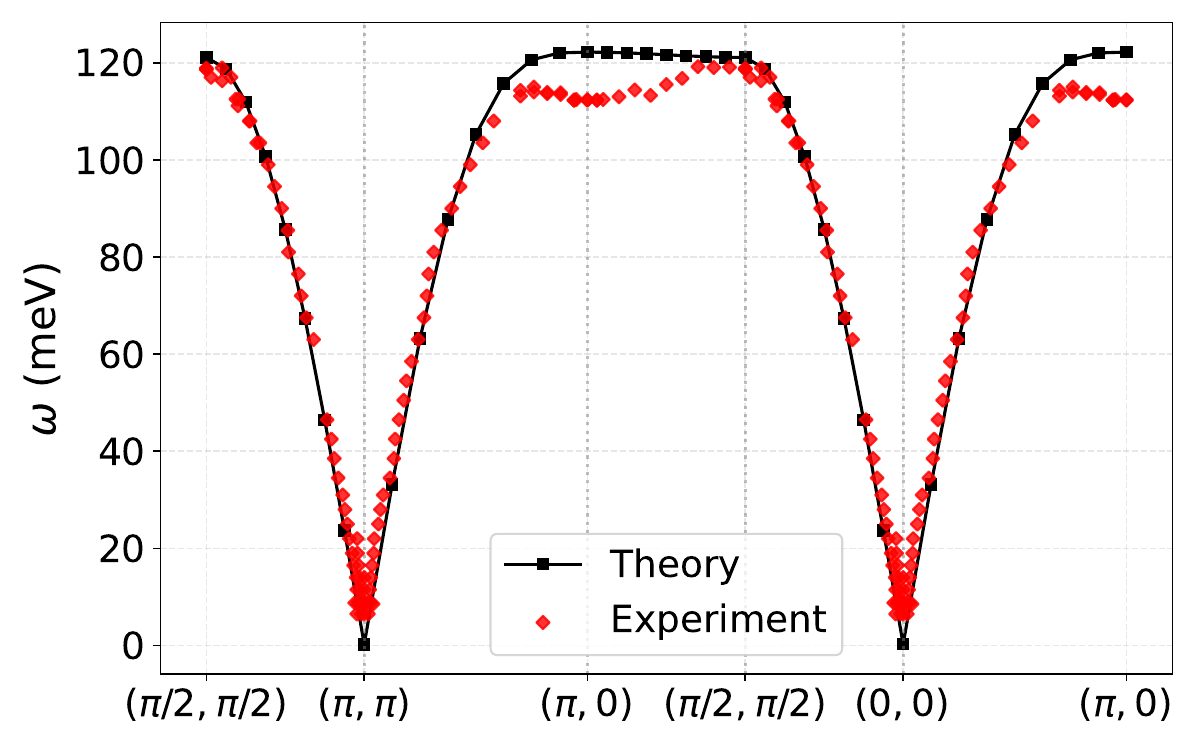}
    \caption{Spin wave dispersion of La$_2$NiO$_4$. In the calculation, the interaction parameters are chosen as $U=6.0\,\mathrm{eV}$ and $J_H=1.8\,\mathrm{eV}$, and the lattice has $64\times 64$ magnetic unit cells. The gap in the experimental spectrum originates from spin orbit coupling, which is not included in the present calculation.}
    \label{fig:TwoOrbDispersion}
\end{figure}

We next apply the method to La$_2$NiO$_4$. 
The low-energy electronic structure of La$_2$NiO$_4$ is represented by a square lattice model with two $e_g$ orbitals and two electrons per Ni site. 
The corresponding tight-binding model for La$_2$NiO$_4$ is provided in the appendix.
The local interaction is taken to be the Kanamori interaction. 
Figure~\ref{fig:TwoOrbDispersion} shows the resulting spin wave dispersion, together with experimental data from inelastic neutron scattering~\cite{petschHighenergySpinWaves2023}.

This calculation demonstrates that the present framework can be implemented in a multi orbital setting relevant to real materials. The low energy part of the calculated dispersion is consistent with the experimental trend. At higher energies, however, the agreement is less quantitative, which may reflect limitations of the simplified two orbital model and the approximate treatment of exchange processes.  The experimental spectrum has a finite gap in low-energy, which may come from spin-orbit coupling. Despite these limitations, the comparison indicates that the method captures the low energy spin dynamics of a correlated multi orbital system and provides a practical starting point for more realistic material calculations.

\section{Conclusion}

In this work, we develop an adiabatic approach for calculating spin wave dispersions in correlated systems by combining the Niu-Kleinman equation of motion with slave boson mean field theory. The central step is to construct a family of constrained saddle point states near the magnetic ground state, from which the Berry curvature matrix and the energy Hessian are extracted. These two quantities determine the linearized adiabatic equation of motion for spin waves.

For the half-filled single-orbital Hubbard model, the NK+KRSB result improves substantially over RPA and agrees much better with DQMC benchmarks, especially near $\vec{q}=(\pi,0)$. The dispersion is also close to the TDGA result, consistent with the connection between KRSB saddle point theory and the Gutzwiller approximation. We further applied the method to a two orbital model of La$_2$NiO$_4$, demonstrating that the same framework can be implemented in a multi-orbital setting relevant to real materials.

The method remains computationally efficient because it only requires constrained saddle point solutions infinitesimally close to the magnetic ground state, and these calculations can be parallelized straightforwardly. Its main limitation is the adiabatic assumption, which is most reliable when low energy spin dynamics are well separated from higher-energy charge and orbital excitations. Within this regime, the combination of Niu-Kleinman dynamics and slave boson theory provides a practical framework for studying spin excitations in strongly correlated systems.

\section{Acknowledgement}
We thank Prof. Xi Dai for valuable discussions and for pointing out the Niu–Kleinman approach.
We acknowledge the support by the National Natural Science Foundation of China (Grant NSFC-12494594), the Chinese Academy of Sciences Project for Young Scientists in Basic Research (2022YSBR-048), the New Cornerstone Investigator Program.

\appendix

\section{Mean Field Hamiltonian in Two Orbital Case}
 
The mean-field Hamiltonian in real space is given by Eq.\ref{HMF_twoorb_realspace}. The local interaction terms and the constraint terms are only function of $\bar\phi$ and independent of spin wave momentum $q$. The most difficult part is the kinetic terms, which contains the renormalization matrix $R$. 

First, we consider the representation of $R$ matrix. It's composed by two parts,
\begin{align}
    R^* &= C M, \\
    C_{\alpha\gamma} &= \sum_{ABnm} \langle A|d_\alpha^\dagger|B\rangle\langle n|f_\gamma^\dagger|m\rangle \phi_{An}^* \phi_{Bm}.
\end{align}
$M$ matrix has been defined in Eq.\ref{M_definition}. With our parameterization of slave bosons, 
\begin{align}
    C_{\alpha,\gamma} &= \sum_{ABnm,A'B'} \langle A|d_\alpha^\dagger|B\rangle\langle n|f_\gamma^\dagger|m\rangle  \notag\\
    &\qquad U_{AA'}^* \bar{\phi}_{A'n}^* U_{BB'} \bar{\phi}_{B'm} \notag \notag\\
    &= \sum_{ABnm,A'B'} \langle A'| \hat{U}^\dagger|A\rangle \langle A|d_\alpha^\dagger|B\rangle \langle B|\hat{U}|B'\rangle \notag\\
    &\qquad \langle n|f_\gamma^\dagger|m\rangle \bar{\phi}_{A'n}^* \bar{\phi}_{B'm} \notag \\
    &= \sum_{A'B'nm}\langle A'|\hat{U}^\dagger d_\alpha^\dagger \hat{U}|B'\rangle \langle n|f_\gamma^\dagger|m\rangle \bar{\phi}_{A'n}^* \bar{\phi}_{B'm} \notag\\
    &= \tilde{u}_{\alpha,\beta} \sum_{A'B'nm} \langle A'|d_\beta^\dagger|B'\rangle \langle n|f_\gamma^\dagger|m\rangle \bar{\phi}_{A'n}^* \bar{\phi}_{B'm} \notag\\
    &= \tilde{u}_{\alpha,\beta} \bar{C}_{\beta,\gamma}.
\end{align}
$\tilde{u}$ is the spin rotation matrix in spin-orbital indices space. It's determined by the spin ratation matrix $U$ written in the local Hilbert space.

$M$ matrix is a function of $\Delta^p$ and $\Delta^h$, which is invariant under the rotation $U$, so
\begin{equation}
    M = \bar{M}.
\end{equation}

The rotation $U$ and $\tilde{u}$ are determined by $S^x$, $S^y$ and $\bar{\phi}$.
\begin{align}
    S_{a} &= \sum_{ABn} \langle A| \hat{S}^z |B\rangle \bar{\phi}_{a,An}^* \bar{\phi}_{a,Bn}, \\
    \sin \theta_{a} &= \frac{\sqrt{(S_a^x)^2 + (S_a^y)^2}}{S_a}, \\
    \tan \varphi_a &= \frac{S_a^y}{S_a^x}, \\
    U_{a,i} &= U_z(\vec{q}\cdot \vec{r}_i) U_z(\varphi_a) U_y(\theta_a), \\
    \tilde{u}_{a,i} &= \tilde{u}_z(\vec{q}\cdot \vec{r}_i) \tilde{u}_z(\varphi_a) \tilde{u}_y(\theta_a).
\end{align}
$a$ is the sublattice index. Substituting these above equations into the kinetic term and do a Fourier transformation, 
\begin{equation}
\begin{aligned}
    H_{kin} =& \sum_{k,\delta} f_{a^0_\delta k}^\dagger \bar{R}_{a^0_\delta}^\dagger \tilde{u}_{a^0_\delta}^T \tilde{u}_z^*(q\cdot\delta) \tilde{u}_{a^1_\delta}^* \bar{R}_{a^1_\delta} f_{a^1_\delta k}, \\
    \tilde{u}_{a} =& \tilde{u}_z(\varphi_a) \tilde{u}_y(\theta_a).
\end{aligned}
\end{equation}
Here the sublattice index $a^0_\delta$ and $a^1_\delta$ are the starting and ending sublattice of hopping $\delta$. $t_\delta$ is hopping matrix which has no spin-flip terms.

The constraint terms is independent of $q$ and spin rotation. It can be directly written in momentum space.
\begin{align}
    H_{con} =& \sum_{k,a=\{a,b\}} \left[f_{ak}^\dagger \Lambda_a f_{ak} - \sum_{\alpha\beta}(\Lambda_a)_{\alpha\beta} (\Delta_a^p)_{\alpha\beta}\right]
\end{align}

\section{Tight-binding Parameters of \texorpdfstring{La$_2$NiO$_4$}{La2NiO4}}

The tight-binding Hamiltonian of La$_2$NiO$_4$ is
\begin{equation}
\begin{aligned}
H ={}& \sum_{\vec{k},\sigma}
\big[
\epsilon_x
+ 2t_{1x}(\cos k_x + \cos k_y)
+ 4t_{2x}\cos k_x \cos k_y \\
+& 2t_{3x}(\cos 2k_x + \cos 2k_y)
\big]
d^\dagger_{\vec{k},x,\sigma} d_{\vec{k},x,\sigma}
\\
+& \sum_{\vec{k},\sigma}
\big[
\epsilon_z
+ 2t_{1z}(\cos k_x + \cos k_y)
+ 4t_{2z}\cos k_x \cos k_y \\
+& 2t_{3z}(\cos 2k_x + \cos 2k_y)
\big]
d^\dagger_{\vec{k},z,\sigma} d_{\vec{k},z,\sigma}
\\
+& \sum_{\vec{k},\sigma}
\big[
2t_{1xz}(\cos k_x - \cos k_y) \\
&\qquad + 2t_{3xz}(\cos 2k_x - \cos 2k_y)
\big] \\
&\left(
d^\dagger_{\vec{k},x,\sigma} d_{\vec{k},z,\sigma}
+
d^\dagger_{\vec{k},z,\sigma} d_{\vec{k},x,\sigma}
\right).
\end{aligned}
\end{equation}

The parameters are in the Table \ref{tbparams_214}.
\begin{table}[htbp]
\centering
\caption{Parameters of the two orbital tight-binding model for La$_2$NiO$_4$. The unit is $eV$.}
\label{tbparams_214}
\begin{tabular}{cccccc}
\toprule
$t_{1x}$ & $t_{1z}$ & $t_{2x}$ & $t_{2z}$ & $t_{3x}$ & $t_{3z}$ \\
\midrule
$-0.4264$ & $-0.076$ & $0.074$ & $-0.0111$ & $-0.0489$ & $0.0082$ \\
\bottomrule
\end{tabular}

\vspace{0.5em}

\begin{tabular}{cccc}
\toprule
$t_{1xz}$ & $t_{3xz}$ & $\epsilon_x$ & $\epsilon_z$ \\
\midrule
$-0.1802$ & $-0.0186$ & $0.4984$ & $-0.0587$ \\
\bottomrule
\end{tabular}

\end{table}

\section{Hartree-Fock Mean Field Theory Combined with Niu-Kleinman Adiabatic Equation}

Before we implement the slave boson mean field theory combined with the Niu-Kleinman adiabatic equation, we first apply the adiabatic equation in Hartree-Fock mean field theory. It's the same as RPA if adiabatic assumption is satisfied. Taking Hubbard model as an example, it starts with the rotation invariant mean-field decoupling
\begin{equation}
\begin{aligned}
    U\left(\hat{n}_{\uparrow}-\frac{1}{2}\right)&\left(\hat{n}_{\downarrow}-\frac{1}{2}\right)=-\frac{2U}{3}\hat{\Vec{S}}\cdot\hat{\Vec{S}}-\mu \hat{n} \\
    &=-\frac{4U}{3}\Vec{m}\cdot\hat{\Vec{S}}-\mu \hat{n}+\frac{2U}{3}\Vec{m}^2
\end{aligned}
\end{equation}
where $\Vec{m}=\langle\Vec{S}\rangle$ is the mean field order parameter, $\mu$ is the chemical potential. The total mean field Hamiltonian is
\begin{equation}
    \begin{aligned}
        H_{MF} =& -t\sum_{\langle ij\rangle,\sigma} \left(c_{i\sigma}^\dagger c_{j\sigma} + h.c.\right) \\
        &+ \sum_i \left( -\frac{4U}{3}\vec{m}_i\cdot \hat{\vec{S}}_i -\mu\hat{n}_i + \frac{2U}{3}\vec{m}_i^2 \right)
    \end{aligned}
\end{equation}

To fix the spin configuration $\Vec{m}$ to a specific value $\vec{m}_{set}$, we can add a large Lagrangian multiplier $A$, e.g. $A=100t$, in the Hartree-Fock mean-field Hamiltonian.\cite{gebauerMagnonsRealMaterials2000}
\begin{equation}
    H_{cons} = H_{MF} + A\sum_i \left( \left\langle\hat{\Vec{S}}_i\right\rangle - \Vec{m}_{set} \right)\hat{\Vec{S}}_i
\end{equation}
One can proove that after add this large multiplier $A$ the spin configuration is fixed at $\Vec{m}_{set}$ while the total mean field energy is the same as before. Now the constraint problem becomes a self consistent problem, which can be easily solved.

\begin{figure}[htbp]
    \centering
    \includegraphics[width=\linewidth]{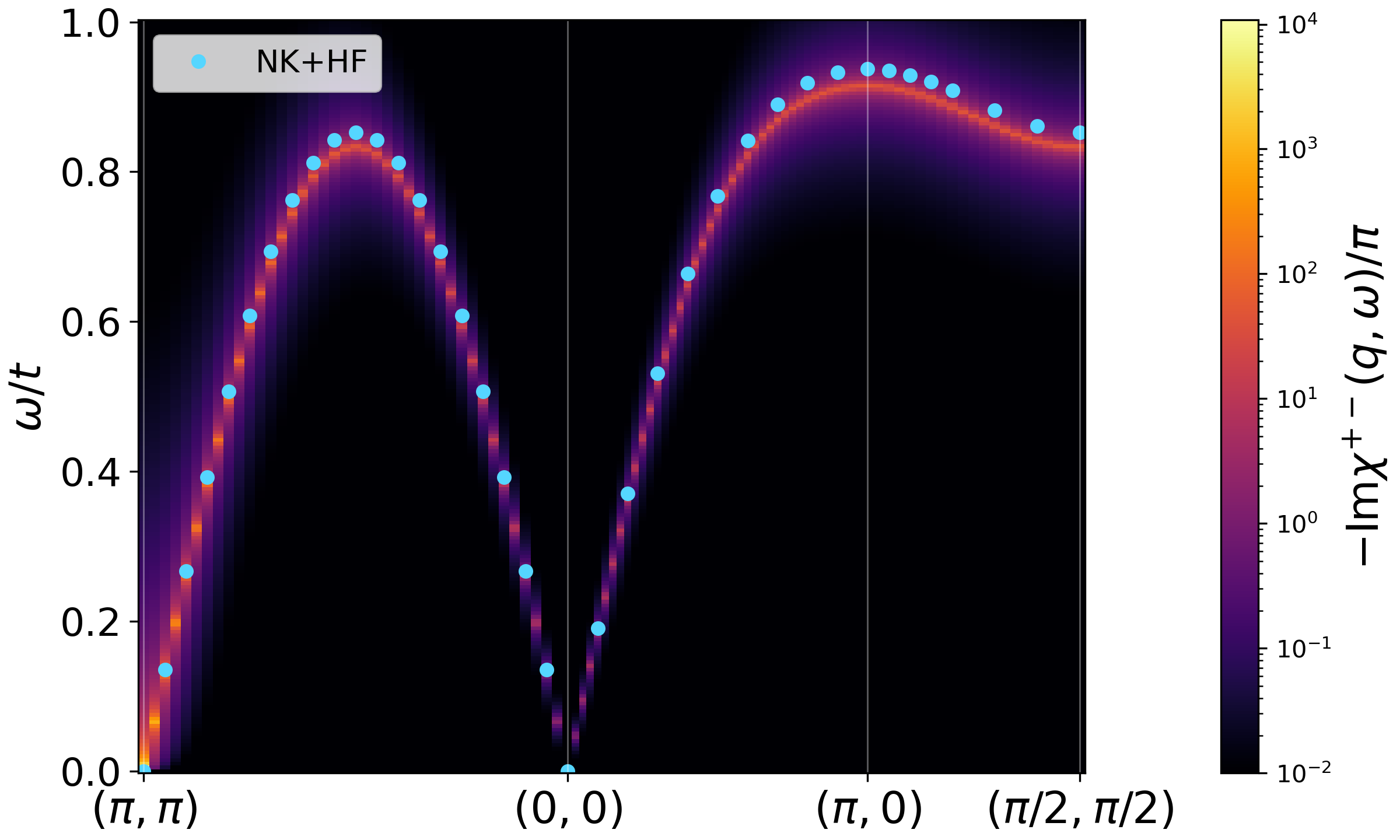}
    \caption{Comparing the result of RPA and NK+HF method. $U=8.0$. The RPA result is almost the same as NK+HF, which indicates that the Niu-Kleinman adiabatic equation captures the essential physics in RPA, at least in the low energy regime.}
    \label{CompareRPA}
\end{figure}

The result Fig.\ref{CompareRPA} shows it agrees well with RPA when $U$ is large. Because the Niu-Kleinman equation uses the adiabatic assumption, the spin wave energy must be in the gap and away from the bottum of other high energy excitations. If U is small, the gap is also small and the spin wave excitation will enter the high energy excitation continuum, which will obviously violate the adiabatic assumption.

\section{Difference of Berry Curvature Calculation Between Single And Two Orbital Case}
\label{DiffOfBerry}

With the parametrization introduced in Section \ref{ParameterizationTwoOrb}, the Berry curvature does not automatically reduce to the single orbital result when the general multi orbital formalism is applied to a single orbital model. The origin of this issue is the gauge convention used for the slave boson fields. In the single orbital case, the slave boson matrix is effectively rotated on both sides,
\begin{align}
    \mathbf{p} &= U \mathbf{p}^{dia} U^\dagger, \\
    Z &= U Z^{dia} U^\dagger.
\end{align}
In the two orbital parametrization used above, however, the spin rotation is applied only to the physical index of $\phi$:
\begin{align}
    \phi &= U \bar{\phi}, \\
    R^\dagger &= \bar{R}^\dagger U^T.
\end{align}
As a result, the kinetic term in $h_k$ differs from the single orbital expression, as can be seen by comparing Eq.~\ref{h_kSingleOrb} and Eq.~\ref{h_kTwoOrb}. In the two orbital expression, the outer rotation matrix appearing in the single orbital form is absent. Although this difference can be viewed as a choice of gauge for the quasiparticle basis, the missing rotation depends on the spin configuration and therefore affects the Berry curvature calculated from the eigenvectors of $h_k$.

To obtain a Berry curvature consistent with the physical spin dynamics, we rotate the eigenvectors of $h_k$ back to the spin frame used in the single orbital convention before evaluating the Berry curvature. This procedure fixes the relevant gauge of the quasiparticle states. The slave boson representation contains a large gauge redundancy, as discussed in Ref.~\cite{lechermannRotationallyInvariantSlaveboson2007}, and the Berry curvature is sensitive to the choice of smooth quasiparticle states along the path in spin configuration space. Since the Berry curvature is computed from the pseudo fermion states, their spin density matrix should be consistent with the physical spin configuration encoded by the slave bosons. This requirement can be written as
\begin{equation}
    \begin{aligned}
        \sum_{ABn} \langle A | d_\alpha^\dagger d_\beta|B\rangle \phi_{An}^* \phi_{Bn} &= \langle f_\alpha ^\dagger f_\beta\rangle \\
        &= \sum_{Anm} \phi_{An}^* \phi_{Am} \langle m| f_\alpha^\dagger f_\beta | n\rangle 
    \end{aligned}\label{ConsistentReq}
\end{equation}
In the convention used in this paper, both $|A\rangle$ and $|n\rangle$ are written in the Fock basis. If the $\phi$ matrix is Hermitian,
\begin{equation}
    \phi^\dagger = \phi,
\end{equation}
then the consistency condition in Eq.~(\ref{ConsistentReq}) is satisfied. Therefore, in the Berry curvature calculation, the rotation matrix should be applied to both sides of $\phi$ and $R$. This choice restores the correct single orbital limit and ensures that the Berry curvature describes the adiabatic motion of the physical spin texture rather than an artifact of the quasiparticle gauge.

One might argue that it is somewhat redundant to discard the rotation matrix in the saddle-point calculation, only to reintroduce it later when evaluating the Berry curvature. However, if we introduce the rotation matrix in the saddle point calculation, it will be hard to write the mean field Hamiltonian in momentum space and the resulting expression will be very complicated.

\section{Berry Curvature of Gaussian states}
\label{BerryCurvOfGauss}
\subsection{Berry Curvature Expressed By Eigen Vectors}
In the main text, the ansatz of $|\Psi[\vec{S}(r)]\rangle$ are constructed from slave boson mean field ground states, which are Gaussian states. The spin wave configuration $\vec{S}(r)$ is controlled by the parameters $x$ in the Hamiltonian $H(x)$, which are slave bosons and Lagrangian multipliers. So the derivatives with respect to $S_j^\alpha$ in Berry curvature can be obtained from derivatives with respect to parameters $x$.

Assume that $H(x)$ is defined as:
\begin{equation}
    H(x) = \sum_{i,j}h_{i,j}(x)c_i^\dagger c_j.
\end{equation}
After diagonalizing $h(x)$, $H(x) = \sum_n\epsilon_n(x) d_n^\dagger d_n$ and $d_n = \sum_i U^*_{in}(x) c_i$.
The ground state is a Fermionic gaussian state, $\left|\Psi(x)\right\rangle = \prod_{\epsilon_n(x)<0} d_n^\dagger(x)|0\rangle$.

First, we know that the derivative of $d_n^\dagger(x)$ is
\begin{equation}
    \begin{aligned}
        \frac{\partial}{\partial x}d_n^\dagger(x) &= \sum_i \frac{\partial U_{in}(x)}{\partial x} c_i^\dagger \\
        &= \sum_{i,n'}\frac{\partial U_{in}(x)}{\partial x} U^*_{in'}(x)d_{n'}^\dagger \\
        &= \sum_{n'} d_{n'}^\dagger \left(U^\dagger\frac{\partial}{\partial x}U\right)_{n'n}.
    \end{aligned}
\end{equation}
Then the first derivative of $|\Psi\rangle$ is (\cite{lenzingEmergentNonAbelianGauge2022})
\begin{equation}
    \begin{aligned}
        &\left\langle\Psi(x)\middle| \frac{\partial}{\partial x}\middle|\Psi(x)\right\rangle  \\
        &= \sum_{i=1}^{N} \Bigl\langle 0\Bigl| \left(U^\dagger\frac{\partial}{\partial x}U\right)_{m,n_i} d_{n_N}\dots d_{n_i}\dots d_{n_1}  \\
        &\qquad d_{n_1}^\dagger\dots d_{m}^\dagger\dots d_{n_N}^\dagger\Bigr| 0\Bigr\rangle \\
        &= \sum_{i=1}^N \left(U^\dagger\frac{\partial}{\partial x}U\right)_{m,n_i} \delta_{n_i,m}= \sum_{\epsilon_m(x)<0} \left(U^\dagger\frac{\partial}{\partial x}U\right)_{m,m}.
    \end{aligned}
\end{equation}
So the Berry curvature w.r.t. parameters $x,y$ is
\begin{equation}
\begin{aligned}
    \Omega_{xy} 
    &= \sum_{\epsilon_m(x)<0} \left(\frac{\partial U^\dagger}{\partial x}\frac{\mathrm{i}\partial U}{\partial y} - \frac{\partial U^\dagger}{\partial y}\frac{\mathrm{i}\partial U}{\partial x}\right)_{m,m} \\
    &= \sum_{\epsilon_m(x)<0}-2\Im\left[ \left(\frac{\partial U^\dagger}{\partial x} U\right)\left(U^\dagger\frac{\partial U}{\partial y}\right) \right]_{mm}
\end{aligned} \label{BerryCurvExpression}
\end{equation}

\subsection{Derivative of Eigen Vectors}
Before using Eq.\ref{BerryCurvExpression} we still need to know how to express the first derivative of eigen vector $u_n(x)$. It's not a trivial task since there's a large gauge redundancy to define a function $u_n(x)$, especially when there's degeneracy.

First let's assume there exists continuous and derivable functions of $x$, $\epsilon_n(x)$ and $u_n(x)$. We have $h u_n=\epsilon_n u_n$ and $u_n^\dagger u_n=1$. Therefore,
\begin{align}
        &\frac{\partial h}{\partial x} u_n + h \frac{\partial u_n}{\partial x} = \frac{\partial \epsilon_n}{\partial x} u_n + \epsilon_n \frac{\partial u_n}{\partial x},
        \label{EigenEqDeriv}  \\
        &\frac{\partial u_n^\dagger}{\partial x} u_n + u_n^\dagger \frac{\partial u_n}{\partial x} = 0. \label{NormEqDeriv}
\end{align}
Then the derivative of eigen value is known if we left multiply $u_n^\dagger$ on both side.
\begin{equation}
    \frac{\partial \epsilon_n}{\partial x} = u_n^\dagger \frac{\partial h}{\partial x}u_n. \label{ResDerivOfE}
\end{equation}
Substituting $\frac{\partial \epsilon_n}{\partial x}$ in Eq.\ref{EigenEqDeriv},
\begin{equation}
    (h - \epsilon_n I)\frac{\partial u_n}{\partial x} = -(I-u_nu_n^\dagger)\frac{\partial h}{\partial x}u_n. \label{FirstDeriveEqOfUn}
\end{equation}

If there's no degeneracy, we know that $I-u_nu_n^\dagger$ project out the components parallel to $u_n$ and $(h-\epsilon_nI)u_n=0$. So the Eq.\ref{FirstDeriveEqOfUn} is possible. If we assume that $\frac{\partial u_n}{\partial x} \perp u_n$, which is reasonable since there's Eq.\ref{NormEqDeriv}, and multiply the pseudo inverse of $H-\epsilon_nI$ on both sides,
\begin{equation}
    \frac{\partial u_n}{\partial x} = -(h-\epsilon_nI)^+\frac{\partial h}{\partial x}u_n. \label{ResDerivOfUn}
\end{equation}
where superscript "$+$" means pseudo inverse of matrix.

If there's degeneracy, it could be proved that Eq.\ref{ResDerivOfE} and Eq.\ref{ResDerivOfUn} are still valid if there's a constraint that $u_n$ is also a eigen vector of $\frac{\partial h}{\partial x}$ in the degenerate subspace. The proof is omitted since the results in this subsection is essentially standard perturbation theory.

\bibliography{refs}

\end{document}